# The Wasserstein Distance for Mortality Comparisons: Absolute versus Net Differences in Survival

Markus Sauerberg[1]

**Abstract**

It is well known that the gap in life expectancy at birth can be seen as the net difference between two survivorship functions. When calculating the absolute difference between the two survivorship functions instead, we derive a distributional inequality measure which is called the Wasserstein distance. The measure quantifies how far apart two probability distributions are by the minimal cost of transporting one into the other. This paper relates the Wasserstein distance to the difference in life expectancy at birth. Both measures correspond to each other, whenever the survivorship functions do not cross, i.e., the net difference equals the absolute difference in the no crossing case. Whether the curves cross is driven by the accumulated difference in age-specific death rates. Crossing occurs in situations where an accumulated survival advantage for given populations is reversed because elevated death rates at later ages outweigh the earlier survival advantage. These cases are particularly interesting because the net difference in survival (or the life expectancy gap) may be small even though the two populations show very different mortality schedules. In our empirical analysis, we search for these cases using life table data from Human Mortality Database life tables for the period 1990 to 2020. Across 69 188 population pairs, survivorship functions cross in 59.8 % of comparisons. Yet, the magnitude of the reversal in survivorship is usually very small. We therefore conclude that in most cases, the gap in life expectancy at birth is not only a comparison of means but quantifies the overall difference between the two mortality regimes.

# Introduction

The Wasserstein distance measures the distance between two probability distributions. It is often demonstrated by imagining each distribution as a pile of dirt. The distance is then the minimal cost of moving one pile into the shape of the other with the cost being the amount of mass moved multiplied by how far it travels. The measure originates in optimal transport theory and can be traced to the problem of finding the optimal transport map between two distributions, first posed by Gaspard Monge in 1781 (Santambrogio 2015; Sauerberg 2025). More recently, the framework has been widely used in the field of computer science (Rubner et al. 2000; Peyré and Cuturi 2019).

Demography has a long-standing interest in examining the difference between two age-at-death distributions. The field has approached it with several measures. Edwards and Tuljapurkar (2005), for example, studied mortality convergence across industrialised countries using the standard deviation of life table ages at death above age ten. Sasson (2016) measured educational differences in American age-at-death distributions with the Kullback–Leibler divergence. Shi et al. (2022) used the non-overlap index, also called the Jaccard index, to compare mortality schedules between Finnish income groups, and Gómez-

[1] Cancer Registry Hamburg, Hamburg, Germany. E-mail: sauerbergmarkus@gmail.com.
Code and replication materials: https://github.com/msauerberg/Wasserstein_repo.

Ugarte et al. (2025) extended the non-overlap index to a multi-group version to study socioeconomic mortality differences in Denmark, England and Sweden. A parallel literature exists for lifespan variation within populations and its relationship to life expectancy (Van Raalte and Caswell 2013; Aburto et al. 2020).

To the best of our knowledge, the first applications of optimal transport in demography are Oeppen et al. (2021), Cilek et al. (2023), and Shang and Haberman (2025). Oeppen et al. (2021) discuss optimisation in life-saving models of mortality, while Cilek et al. (2023) use the Wasserstein distance to study cause-specific mortality differences between border regions. Shang and Haberman (2025) forecast age-at-death distributions using a cumulative distribution function transformation that is closely related to the Wasserstein distance as we show later in this paper.

The conceptual difference between the Wasserstein distance and previously used dissimilarity measures is that it defines statistical distance in terms of *transporting mass*. The choice of how transport cost is measured is therefore key and it determines the demographic interpretation. When deaths are compared on the age axis alone and the cost of moving a death from one age to another is the absolute difference in years, the distance is measured in years. In this case, the Wasserstein distance is closely related to the difference in life expectancy at birth.

This paper has five aims. First, we introduce the Wasserstein distance to demography and establish formally when it coincides with the difference in life expectancy at birth. Second, we ask *when and why* survivorship functions cross, expressing the condition in terms of age-specific death rates through the function $Z(x)$, and we quantify how often crossings occur, depending on the underlying population (life table population, real population or constant-birth population model). Further, we derive from two additional measures, the reversal share $\Lambda$, which is the ratio of the gap in life expectancy at birth and the Wasserstein distance, and the effective number of advantage regimes $\Psi$, which is the inverse Simpson index and records how the survival advantage is distributed over age. Together with the gap in life expectancy at birth and Wasserstein distance, these give a compact description of a mortality difference: net difference, absolute difference, and how much of the difference reverses with age. Third, we explain analytically and verify empirically why the Wasserstein distance correlates so strongly with the non-overlap index. Fourth, we describe two decomposition methods for the one-dimensional Wasserstein distance. Fifth, we extend the Wasserstein distance to a two-dimensional measure that allows measuring distributional differences in cause-specific mortality. Throughout, the empirical analysis is based on data from the Human Mortality Database (Human Mortality Database 2026).

# Methods

## The Wasserstein distance

In the Kantorovich formulation, the distance between two probability distributions $P$ and $Q$ is

$$W_p(P,Q) = \left( \inf_{J \in \mathcal{J}(P,Q)} \int \| x - y \|^p \ dJ(x,y) \right)^{1/p}, \qquad (1)$$

where $\mathcal{J}(P,Q)$ is the set of all couplings, or transport plans, between $P$ and $Q$. The infimum selects the cheapest plan. For each pair $(x,y)$, the cost of moving mass from $x$ to $y$ is $\| x - y \|^p$ and the outer exponent $1/p$ returns the result to the units of the underlying metric. For $p = 1$ the measure is called the Earth Mover's distance, the minimum average absolute distance mass must travel to transform $P$ into $Q$ (Wasserman 2019).

Equation (1) is a general formulation, but in one dimension the transport problem is far simpler. Let $F_P$ and $F_Q$ be the cumulative distribution functions of $P$ and $Q$, with quantile functions $F_P^{-1}$ and $F_Q^{-1}$. Then

$$W_p(P,Q) = \left( \int_0^1 \left| F_P^{-1}(u) - F_Q^{-1}(u) \right|^p du \right)^{1/p}, \qquad (2)$$

and in the special case $p = 1$ the distance also equals the area between the cumulative distribution functions,

$$W_1(P,Q) = \int_{-\infty}^{+\infty} \left| F_P(x) - F_Q(x) \right| dx \qquad (3)$$

see Santambrogio 2015, chap. 2 for more details. Equation (3) is used to link $W_1$ to life expectancy at birth differentials.

Using $p = 1$ treats transport cost linearly, i.e., moving mass from age ten to age zero costs $|10 - 0| = 10$, so the distance is measured simply as a difference in years, and $W_1$ is itself expressed in years. With $p = 2$ the cost grows quadratically, $(|10 - 0|)^2 = 100$, which makes long transports disproportionately expensive and removes the direct reading of the result as an age difference. For comparing age-at-death distributions the linear cost is usually the more meaningful choice, and it is the one that connects the measure to life expectancy at birth differentials.

## $W_1$ equals the gap in life expectancy at birth when survivorship functions do not cross

Let $d_A$ and $d_B$ denote two life table age-at-death distributions with survivorship functions $l_A$ and $l_B$. We set the radix to one, $l_A(0) = l_B(0) = 1$, so that $d(x)$ is a probability density and $l(x)$ is the corresponding survival function, $F(x) = 1 - l(x)$. The probability of dying in the last open age interval is one, $l_A(\omega) = l_B(\omega) = 0$.

**Result 1**. *If the survivorship functions do not cross, that is if*

$$l_A(x) \geq l_B(x) \quad \forall x \in [0, \omega], \qquad (4)$$

*Then*

$$W_1(d_A, d_B) = e_{0,A} - e_{0,B}. \qquad (5)$$

The argument is straight forward. Since $F_A = 1 - l_A$ and $F_B = 1 - l_B$, the integrand of (3) is $|l_A(x) - l_B(x)|$. Whenever (4) holds it is non-negative, so the absolute value can be dropped and the integral becomes $\int_0^{\omega}[\, l_A(x) - l_B(x)]\, dx$, which is precisely the gap in life expectancy at birth, $e_{0,A} - e_{0,B}$. Appendix A gives the derivation in full, together with the standard equivalence between life expectancy expressed as the area under the survivorship function and as the mean of the age-at-death distribution. The latter indicates that Result 1 can be read as a statement about means, i.e., when the survivorship functions do not cross, the Wasserstein distance between two age-at-death distributions equals the difference between their mean ages at death. In other words, for age-at-death distributions satisfying (4), the difference in means *is* the full optimal transport cost, so the gap in $e_0$ can be interpreted directly as a distributional distance and not only as a comparison of averages.

The difference between life expectancy at birth values can be seen as the net difference between the two corresponding survivorship functions, $\Delta(x) = l_A(x) - l_B(x)$. The positive and negatives parts can be written as,

$$P = \int_0^{\omega} \max\{\Delta(x),0\}\, dx, \qquad N = \int_0^{\omega} \max\{-\Delta(x),0\}\, dx, \qquad (6)$$

with $P$ and $N$ being the areas over which $A$ survives better or worse than $B$, respectively. The *P* term is simply the sum of the positive differences between $l_A(x)$ and $l_B(x)$ over age, while *N* gives the sum of all negative age-specific differences between the two functions. Hence, the net difference in survival or the life expectancy at birth differential can be written as, $\Delta e_0 = P - N$. Accordingly, the absolute difference in survival, which is equivalent to $W_1$ is given by $W_1 = P + N$. Assuming that $A$ is the population with higher life expectancy at birth compared to $B$, we can write the difference between the absolute and net difference in survival as, $W_1 - \Delta e_0 = (P + N) - (P - N) = 2N$. Note that this only holds in the case of a higher $e_0$ value for population *A*. For the general case, we can write $W_1 - |\Delta e_0| = 2\min(P, N)$, which is independent of the choice $\Delta e_0 = l_A(x) - l_B(x)$ vs. $\Delta e_0 = l_B(x) - l_A(x)$. This implies directly,

$$W_1(d_A, d_B) = |\Delta e_0| + 2\min(P, N). \quad (7)$$

The term $2\min(P, N)$ is exactly the amount by which the Wasserstein distance exceeds the life expectancy gap whenever (4) does not hold. If there is no crossing between $l_A(x)$ and $l_B(x)$, the term $\min(P, N)$ equals zero and $W_1$ corresponds to $\Delta e_0$ exactly. If there is, however, a crossover, $W_1$ is given by adding the $e_0$ differential to the double of the smaller area *P* or *N*. Equation (7) turns Result 1 from a conditional statement into a general identity that holds for every pair. Two immediate implications are that $W_1 \geq |\Delta e_0|$ always, and that a pair with $\Delta e_0 = 0$ can still show a large $W_1$ value.

When $W_1$ and $\Delta e_0$ do not correspond to each other, it makes sense to look at the size of $2\min(P, N)$, which is precisely the amount of absolute difference in survival that was canceled out in net survival differences. We define the measure $\Lambda$ as the share of the absolute difference in survival that is offsetting rather than net,

$$\Lambda = \frac{2\min(P, N)}{W_1} = 1 - \frac{|\Delta e_0|}{W_1} \in [0,1]. \qquad (8)$$

The measure is zero when one population's survivorship function dominates at every age, and one when net survival is very similar (i.e., the two life expectancies values are similar) despite a large $W_1$. The three quantities then read as a set: $\Delta e_0$ measures net survival, $W_1$ measures absolute survival, and $\Lambda$ measures how much of the absolute difference is advantage cancelling against disadvantage, which is a useful quantity that decides whether a life expectancy gap is an adequate summary of a mortality difference.

## When and why do survivorship functions cross?

Condition (4) is stated in terms of $l(x)$, but it is useful to express it in terms of age-specific death rates. The two are linked through the cumulative hazard. Let $\mu(x)$ be the force of mortality, then $l(x) = \exp\{-\int_0^x \mu(a)\, da\} = \exp\{-H(x)\}$. We can define,

$$Z(x) = \int_0^x [\mu_A(t) - \mu_B(t)]\, dt = H_A(x) - H_B(x) = -\ln\frac{l_A(x)}{l_B(x)}. \qquad (9)$$

In discrete terms, we can express the ratio of survivorship functions in terms accumulated age-specific survival probabilities, $R(x) = l_A(x)/l_B(x) = \prod_{i=0}^{x-1} p_i^A / p_i^B$. Accordingly, $Z(x) = -\ln R(x)$ and reflects the accumulation of the log ratio of age-specific survival probabilities.

The function $Z(x)$ helps us understand when survivorship functions cross. First, $Z(x) < 0$ exactly where $l_A(x) > l_B(x)$, so (4) holds if and only if $Z(x) \leq 0$ throughout, and every sign change of $Z$(x) is a crossover of the two survivorship functions. Second, $Z(0) = 0$ by construction, so the question is whether $Z$ ever returns to zero after leaving it. Third, the derivative is

$$Z'(x) = \mu_A(x) - \mu_B(x), \qquad (10)$$

the difference in age-specific death rates at age $x$. $Z$ therefore falls at ages where population $A$ has the lower death rate and rises where it has the higher one.

A crossing at age $x^*$ requires $Z(x^*) = 0$ with $Z$ changing sign there, which by (10) requires two things. The age-specific death rates must themselves cross at some earlier age. This is necessary but not sufficient. In addition, the accumulated excess mortality of $A$ after the rate crossing must be large enough to cancel its accumulated deficit before. Crossings of survivorship functions are thus rarer than crossings of death rates. They occur later because the rates must diverge in the opposite direction for a long enough time and at ages that carry enough mortality for the accumulated advantage to be erased.

One special case should be noted explicitly as it is widely used in mortality models. Suppose the two mortality schedules are proportional,

$$\mu_A(x) = \theta\, \mu_B(x) \qquad \text{for all } x, \qquad (11)$$

with a constant hazard ratio $\theta > 0$. Substituting (11) into (9),

$$Z(x) = \int_0^x [\theta\, \mu_B(t) - \mu_B(t)]\, dt = (\theta - 1)\, H_B(x), \qquad (12)$$

implies that $Z(x)$ never returns to zero and thus, proportional hazards cannot produce a crossover in survivorship functions. This has direct implications for several demographic applications as we mention in the Discussion section of this paper.

## Does the underlying population model play a role for the observed number of crossings?

The formal analysis of survivorship function crossovers so far refers to life table age-at-death distributions exclusively. In this framework, the deaths of a synthetic cohort are exposed to one period's death rates at every age. Demographers, however, may want to compare the age-specific deaths that population actually observed or the age-at-death distribution implied by the constant-birth population. The latter has been discussed in context of the alternative mortality measure cross-sectional average length of life (CAL) in the previous literature (Brouard 1986; Guillot 2003). It is worthwhile comparing these frameworks because all of them use the same set of period death rates. Thus, differences in the age-at-death distributions can be fully attributed to differences in the underlying population, i.e., the stationary life table population, the real population, and the constant-birth population (Wilmoth 2005; Sauerberg & Luy 2025). Both, the stationary life table population and the constant-birth population are solely derived from of age-specific death rates and therefore, highly unrealistic population age structures. To assess whether the number of crossings is partly the result of using an unrealistic population age structure, we define equivalent $W_1$ results on the basis of the real population as well as the constant-birth population.

The life table age-at-death distribution age $x$ is given by $d(x) = l(x)\, \mu(x)$, whereas the age-at-death distribution of the real population (exposed to changes in fertility, mortality, and migration) is given by, $D^{\text{obs}}(x) = N(x)\, \mu(x)$, with $N(x)$ being the observed population at age $x$. Normalising these counts gives $f(x) = N(x)\mu(x) / \sum_t N\,(t)\mu(t)$ and the complementary distribution function

$$\bar{G}(x) = \Pr(\text{age at death} \geq x) = 1 - \sum_{t<x} f\,(t). \;\; (13)$$

The function $\bar{G}$ is not a survivorship function because it is shaped by the age structure of the real population as much as by its mortality. Nevertheless, we can derive the Wasserstein distance using $W_1 = \int |\bar{G_A}(x) - \bar{G_B}(x)|$ and the difference in mean ages at death is given by $\int (\bar{G_A}(x) - \bar{G_B}(x))$. Hence, the identity (7) continues to hold with $\Delta e_0$ replaced by the gap in mean age at death, and $\min(P, N) = 0$ if and only if the two curves never cross.

The constant-birth population model considers a population that is subjected to a constant stream of births and is closed to migration. A person aged $x$ in year $t$ was born in $t - x$, so the proportion of that birth cohort still alive is the product of period survival probabilities along the Lexis diagonal,

$$S(x,t) = \int_0^{\omega} p\,(x,\ t-x) = \exp\left\{-\int_0^{x} \mu\,(a,\ t-x)\,da\right\}, (14)$$

To analyze whether crossings become more frequent outside of the life table framework we can look at the impact of the three population weights, $l(x)$, $N(x)$, $S(x)$. For instance, we can express $D^{obs}(x)$ in terms of the life table age-at-death distribution, $d(x)$, reweighted by the ratio of the observed to the stationary age structure,

$$D^{\text{obs}}(x)\ \propto\ d(x)\,\frac{N(x)}{l(x)}.\ (15)$$

The reweighting factor $N(x)/l(x)$ shows the large difference between the real population and the life table population. While $l(x)$ is a smooth and monotone decreasing function the function $N(x)$ may rise and fall with age. In contrast to $l(x)$, it is not only a function of current death rates but shaped by past fertility, mortality, and migration. To compare two populations in terms of their age-at-death distributions we can write,

$$\frac{D_A^{obs}(x)}{D_B^{obs}(x)} = \frac{\mu_A(x)}{\mu_B(x)}\,\frac{N_A(x)}{N_B(x)}, \qquad (16)$$

the product of a mortality ratio and an age-structure ratio. Analogously, the ratio, $d_A(x)/d_B(x)$, is given by $\frac{\mu_A(x)}{\mu_B(x)}\,\frac{l_A(x)}{l_B(x)}$. The ratio $l_A(x)/l_B(x)$ varies far less with age than the age-structure ratio, $N_A(x)/N_B(x)$. Further, the ratio of the population weights determines at which ages the two densities differ and more mass (or deaths) lead to a steeper-falling survivorship function. At these ages, the two curves can cross and these additional crosses can be attributed to the difference in the population weights. Judging from the shape of the population weights, one might expect to observe the highest number of crossings using $N(x)$ and the lowest number on the basis of $l(x)$. The constant-birth population standardizes for changes in fertility and assumes no migration and its number of crossings is expected to lie between the other two.

## Introducing a magnitude-weighted measure for survivorship curve crossings: The effective number of advantage regimes $\Psi$

A crossing is simply a change of sign and a sign change carries no information about magnitude. For instance, two curves that graze one another at age 104 register the same "one crossing" as two curves that swap the advantage at age 60 and split the area between them evenly. The sign pattern of $Z(x)$ supports two magnitude-weighted measures that do not have this defect.

$Z(x)$ can be seen as a function that partitions the age range into the maximal length of constant signs. Within a constant sign portion the same population holds the survival advantage throughout, and each sign change hands the advantage over. We call these portions *advantage regimes* and index them by $r = 1, \dots, k$, so that $k$ equals the number of crossings plus one. Each regime is worth what it commands of the area between the curves:

$$m_r = \sum_{x \in \text{regime } r} |l_A(x) - l_B(x)|, \qquad \sum_{r=1}^{k} m_r = W_1, \qquad \theta_r = \frac{m_r}{W_1}. \qquad (17)$$

The regime-specific masses, $m_r$, are in years and split $W_1$ exactly. Hence, $W_1$ is given by summing the masses over all advantage regimes. Calculating the regime-specific weights, $\theta_r$, enables us to quantify how much mass is actually carried by a given regime. These weights can then be used to construct the inverse Simpson index of the regime masses,

$$\Psi = \frac{1}{\sum_{r=1}^{k} \theta_r^2} \in [1, k]. \qquad (18)$$

$\Psi$ is the number of equally weighted regimes that would produce the same concentration of area. It equals one when a single regime carries everything and $k$ when all $k$ regimes command an equal share (Simpson 1949; Hill 1973).

The two measures, $\Lambda$ and $\Psi$, are related but not redundant. Consider two comparisons with regime masses $(2.10, 0.90)$ and $(1.80, 0.90, 0.30)$, the middle regime favouring B in both. They agree on $W_1 = 3.00$, on $\Delta e_0 = 1.20$ and hence on $\Lambda = 1 - \frac{1.20}{3.00} = 0.60$, but $\Psi$ is 1.72 in the first and 2.17 in the second. Because $\Lambda$ is simply putting the term $2\min(P, N)$ into perspective its contribution is interpretive. In contrast, $\Psi$ provides new information on diversity in mass of the regimes and hence, is a genuinely additional statistic.

## Relationship between the $W_1$ Wasserstein distance and the non-overlap index

As mentioned in the Introduction, Shi et al. (2022) introduced the non-overlap index, based on the Jaccard similarity

$$J(p, q) = \frac{\int \min\{p(x), q(x)\}\, dx}{\int \max\{p(x), q(x)\}\, dx}, \qquad (19)$$

which equals one for two identical distributions and zero for two non-overlapping distributions. Consequently, the non-overlap index is defined as $1 - J$. The conceptual difference between $W_1$ and $J$ is that $W_1$ measures how far mass must move, while $J$ measures how much mass is shared. Yet, the correlation between both measures when applied to age-at-death distributions is strong (see Results section). The relationship can be examined formally.

Following Moulton and Jiang (2018), $J$ is an exact function of the total variation distance, $J = (1 - \mathrm{TV})/(1 + \mathrm{TV})$ with $\mathrm{TV}(p, q) = 1/2 \int |p - q|$. Now suppose the two age-at-death distributions differ by a pure location shift, $q(x) = p(x - \delta)$. A first-order expansion gives $\mathrm{TV}(\delta) \approx \delta/2 \int |p'(x)| dx$, and for any unimodal density $\int |p'(x)| dx = 2p(M)$ with $M$ the modal age at death, so $\mathrm{TV} \approx \delta\, p(M)$. Under a pure shift $W_1 = \delta$ exactly, and therefore

$$J \approx \frac{1 - W_1\, p(M)}{1 + W_1\, p(M)}. \qquad (20)$$

See Appendix for more details. Equation (20) reveals that we can express $J$ as a function of $W_1$ under the pure location shift assumption. Given that age-at-death distributions in high-income countries usually share a somewhat typical shape, the expression may provide a mathematical explanation for the high correlation between $J$ and $W_1$.

## Extension to cause of death

The $W_1$ measure can be extended to cause-specific mortality (Cilek et al. 2023). Multi-decrement life tables yield age-at-death distributions by cause (Preston et al. 2001). When using the radix of one, summing $d(x, c)$ over ages $x$ and causes $c$ gives one, making the joint distribution over age and cause a proper probability distribution. While the one-dimensional Wasserstein distance has the closed form (3), the two-dimensional transport problem needs to be solved numerically. We use the network simplex algorithm implemented in the Python Optimal Transport package to find the optimal transport plan (Flamary et al. 2021). Similar results can be derived in R as well when using the “transport” package.

Two dimensions require a cost of moving mass between age and cause of death,

$$M_{ij} = |x_i - x_j| + \lambda \mathbf{1}\{c_i \neq c_j\}, \qquad (21)$$

where the first term is the transport cost in years of age and $\lambda$ is the cost of reallocating a death between causes. The choice of $\lambda$ is a substantive one and it has no natural scale. Setting $\lambda = 0$ makes movement between causes free, so the measure reduces to a comparison of age patterns. When setting $\lambda \to \infty$, no transports between causes of death are allowed, so the problem separates into independent per-cause transports. Further, setting $\lambda = 1$ declares reallocating a death across causes to be as costly as moving it one year of age. It is possible to use different lambdas for different causes of death. For instance, a transport from external causes to cancer can be more costly than a transport from cancer to cardiovascular diseases. A pragmatic strategy might be anchoring $\lambda$ to the all-cause one-dimensional distance, i.e., the $W_1$ Wasserstein distance. In our empirical analysis, we report results for several values to demonstrate that the choice of lambda matters.

## Decomposing the Wasserstein distance

Changes in the age-at-death distribution might reflect shifts to older ages, i.e., the distribution moves on the x-axis to older ages due to mortality improvements or the shape of the distribution changes due to a compression of mortality, i.e., a larger proportion of deaths fall in a small age range. To distinguish between these two scenarios, it is helpful to decompose the Wasserstein distance into shift and shape components.

Recently, Resin et al. (2025) introduced a decomposition method for the $W_1$ Wasserstein distance. Their decomposition method is based on the quantile function representation of the $W_1$ Wasserstein distance (equation 2). This allows expressing $W_1$ as four additive terms,

$$W_1(\mathrm{P}, \mathrm{Q}) = \mathrm{Shift}_+(P, Q) + \mathrm{Shift}_-(P, Q) + \mathrm{Disp}_+(P, Q) + \mathrm{Disp}_-(P, Q), \qquad (22)$$

where the positive shifting term captures an upward shift of $P$ relative to $Q$, the negative shifting term reflects a downward shift of $P$ relative to $Q$, the positive dispersion term quantifies the extent to which $P$ is more dispersed than $Q$, and the negative dispersion term reflects the extent to which $P$ is less dispersed than $Q$. For more details on the decomposition method refer to Resin et al. (2025) or to Appendix.

The second decomposition applies to the squared second-order distance, $\mathrm{W}_2(\mathrm{P},\mathrm{Q}) = \int_0^1 | F_P^{-1}(u) - F_Q^{-1}(u)|^2 du$, and splits it into three terms (Irpino and Verde 2015; Schefzik et al. 2021),

$$\mathrm{W}_2(\mathrm{P},\mathrm{Q}) = \underbrace{(\mu_P - \mu_Q)^2}_{\text{location}} + \underbrace{(\sigma_P - \sigma_Q)^2}_{\text{size}} + \underbrace{2\sigma_P\sigma_Q(1-\rho)}_{\text{shape}}, \qquad (23)$$

where $\mu$ and $\sigma$ are the mean and standard deviation of the age-at-death distribution and $\rho$ is the Pearson correlation between the points of the quantile-quantile plot of the two distributions. Location is the gap in mean age at death, size the gap in its dispersion, and shape whatever remains once means and spreads have been matched. The shape term vanishes only if one distribution is an affine transformation of the other.

The two decompositions are not comparable term by term and are not intended to be. While the first apportions the Wasserstein distance in the unit of years, the second uses quadratic terms, making those years of discrepancy that are at higher ages count for much more than a year near the middle.

# Data

The analysis uses period life tables, death counts, and exposures from the Human Mortality Database (Human Mortality Database 2026). All 50 population codes are used, with women and men treated separately throughout. Three comparisons are made. The cross-country comparison takes all pairs of populations within the same calendar year and sex over 1990 to 2020, giving 69 188 comparisons. The sex gap comparison takes women against men within the same population and year, giving 1 479 comparisons. The time trend comparison takes every combination of a population in 1990 with a population in 2019, by sex, giving 4 136 comparisons. The 82 of those that compare a population with itself are reported separately as the pure within-population trend.

# Results

## Graphical illustration on the basis of two examples

The top row of Figure 1 compares Japanese and United States women in 2019. The Japanese age-at-death distribution is located to the right of the American one at essentially every age, the survivorship functions never touch, $Z(x)$ is negative throughout and never changes sign, and consequently $W_1$ and $\Delta e_0$ coincide at 5.84 years. The entire distributional difference is captured by the gap in life expectancy. The bottom row compares England & Wales with Iceland in 1849, two populations whose life expectancies are indistinguishable ($\Delta e_0 = 0.02$ years). Iceland had much higher infant and early-childhood mortality, so its survivorship function starts below that of England & Wales, but its mortality was lower from mid-life onwards, and by age 44 the accumulated advantage has reversed the sign of $Z(x)$. The positive and negative areas nearly cancel, which is why the life expectancy gap is almost zero, but they do not cancel in absolute value ($W_1 = 6.20$ years) of which $2\min(P, N) = 6.18$ is not included in $\Delta e_0$. This corresponds to $\Lambda = 0.997$ with $\Psi = 2.00$. Two advantage regimes of almost exactly equal weight and a distance that is very nearly all reversal. This is the case in which reporting only the difference in life expectancy would mislead.

**Fig. 1 The Wasserstein distance, the life expectancy gap, and the crossover indicator Z(x)**

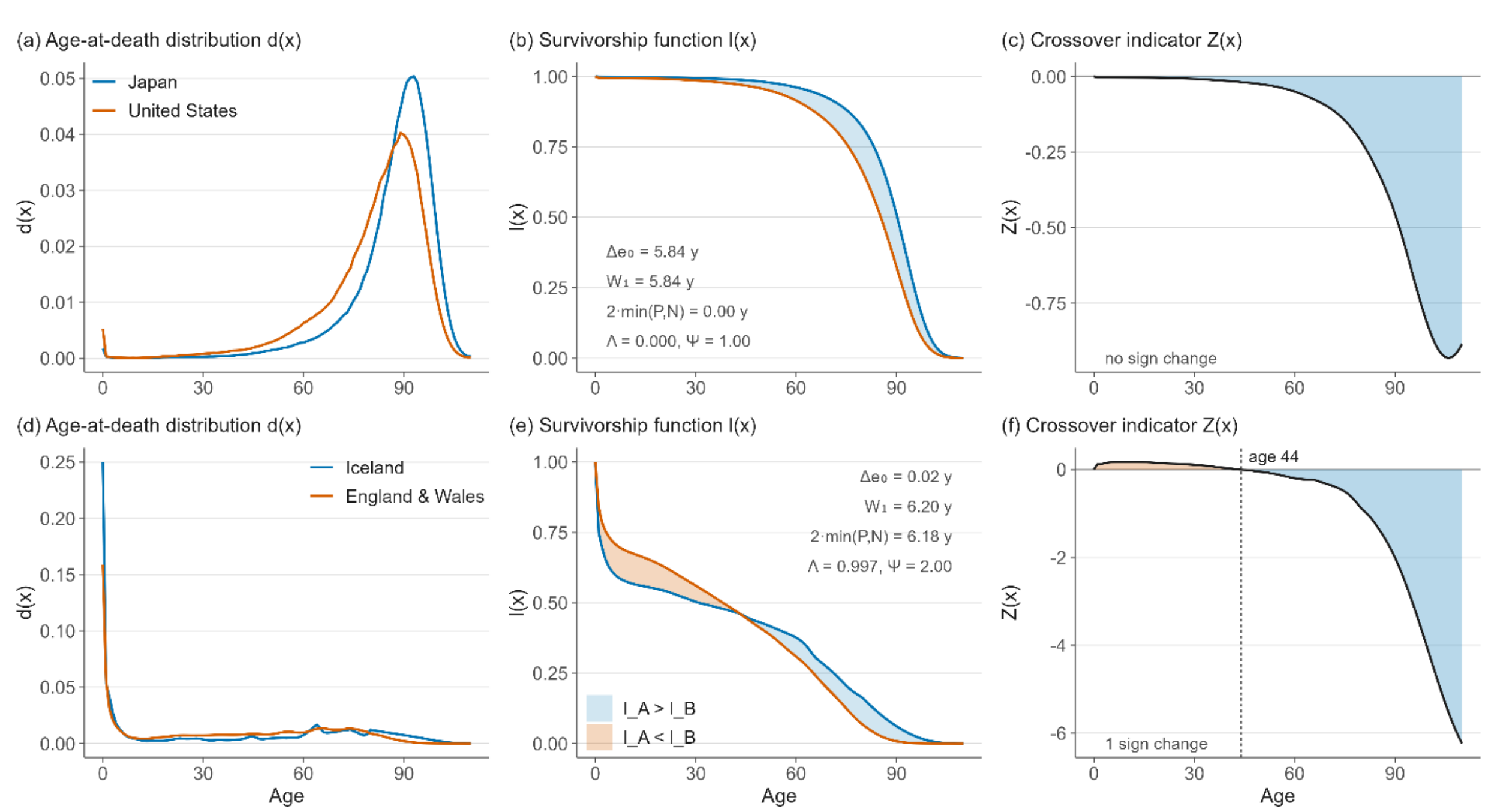

Source: *Own calculations from Human Mortality Database (2026).*

## How often, and at what ages, do survivorship functions cross?

Tables 1 and 2 answer the empirical question for the three comparisons. The results suggest that how often the curves cross depends strongly on what is being compared.

**Cross-country:** Among the 69 188 comparisons of two populations in the same year and sex, the survivorship functions cross at least once in 59.8% of cases and more than once in 27.4%. The textbook picture in which one population dominates the other at every age therefore describes 40% of comparisons. Further, $\Lambda > 0.05$ indicates in how many cases the gap in life expectancy at birth captures less than 95% of the absolute difference in the underlying survivorship functions. This is in 30% of all 69,188 cross-country pairs. The distribution of ages when first crossing occurs is spread out with a median of 45 and highest densities observed at young and old ages (see Figure 2). When considering no mortality before age 30, the median increases to 49. We observe a spike at around age 30. This makes sense because it is more likely to reverse an advantage in survival at the beginning of the accumulation process, i.e., when $Z(x)$ does not deviate from zero very much yet. The spike is more pronounced for the "no mortality before age 30" scenario because death rates are extremely low at very young ages.

**Table 1 Relationship between $W_1$Wasserstein distance and differences in life expectancy at birth: Three comparisons, all ages**

| | | Crossings | | | | Years | | | Weighed | |
|---|---|---|---|---|---|---|---|---|---|---|
| | $n$ | $\geq 1$ (%) | $> 1$ (%) | Median first age | $\Lambda > 0.05$ (%) | Mean $\Delta e_0$ | Mean $W_1$ | Mean $2\min(P,N)$ | Mean $\Lambda$ | Mean $\Psi$ |
| Cross-country | 69,188 | 59.8 | 27.4 | 45 | 30.0 | 3.57 | 3.69 | 0.118 | 0.108 | 1.165 |
| Time trend, 1990 vs. 2019 | 4,136 | 27.8 | 8.7 | 100 | 9.2 | 6.58 | 6.64 | 0.060 | 0.036 | 1.072 |
| same population | 82 | 7.3 | 0.0 | 105 | 0.0 | 6.15 | 6.15 | 0.0001 | 0.0000 | 1.000 |
| Sex gap | 1,479 | 10.2 | 3.8 | 12 | 0.0 | 6.30 | 6.30 | 0.001 | 0.0002 | 1.000 |

Source: *Own calculations from Human Mortality Database (2026).*

**Table 2 Relationship between $W_1$Wasserstein distance and differences in life expectancy at birth: Three comparisons, no mortality before age 30**

| | | Crossings | | | | Years | | | Weighed | |
|---|---|---|---|---|---|---|---|---|---|---|
| | $n$ | $\geq 1$ (%) | $> 1$ (%) | Median first age | $\Lambda > 0.05$ (%) | Mean $\Delta e_0$ | Mean $W_1$ | Mean $2\min(P,N)$ | Mean $\Lambda$ | Mean $\Psi$ |
| Cross-country | 69,188 | 58.9 | 26.4 | 49 | 23.5 | 3.25 | 3.34 | 0.080 | 0.091 | 1.130 |
| Time trend, 1990 vs. 2019 | 4,136 | 36.6 | 13.2 | 60 | 9.2 | 5.72 | 5.78 | 0.049 | 0.035 | 1.048 |
| same population | 82 | 18.3 | 9.8 | 34 | 0.0 | 5.30 | 5.30 | 0.003 | 0.0006 | 1.001 |
| Sex gap | 1,479 | 4.3 | 0.7 | 103 | 0.1 | 5.78 | 5.78 | 0.0002 | 0.0001 | 1.000 |

Source: *Own calculations from Human Mortality Database (2026).*

**Time trend:** Comparing populations in 1990 with populations in 2019 gives a very different picture. Crossings occur in 27.8% of the 4 136 combinations, and more than once in only 8.7%. Restricting to the 82 pairs that compare a population with itself 29 years later, in the pure within-population trend just 7.3% cross and none of them more than once. The median first crossing falls at age 105. Mortality improvement between 1990 and 2019 seems to be smooth, meaning that the survivorship curve increases at nearly every age. There are no meaningful crossings so that the mean $2\min(P,N)$ of 0.0001 years makes $W_1$ and $\Delta e_0$ interchangeable for measuring progress over this period.

**Sex gap:** Comparing women with men within the same population and year, crossings occur in 10.2% of the 1 479 country-years and the mean discrepancy is 0.001. Female survivorship dominates male survivorship at essentially every age, making the Wasserstein distance and the life expectancy gap the same quantity to two decimal places.

**Fig. 2 Number of crossings and age at the first crossing, cross-country comparisons, life table framework, 1990 to 2020**

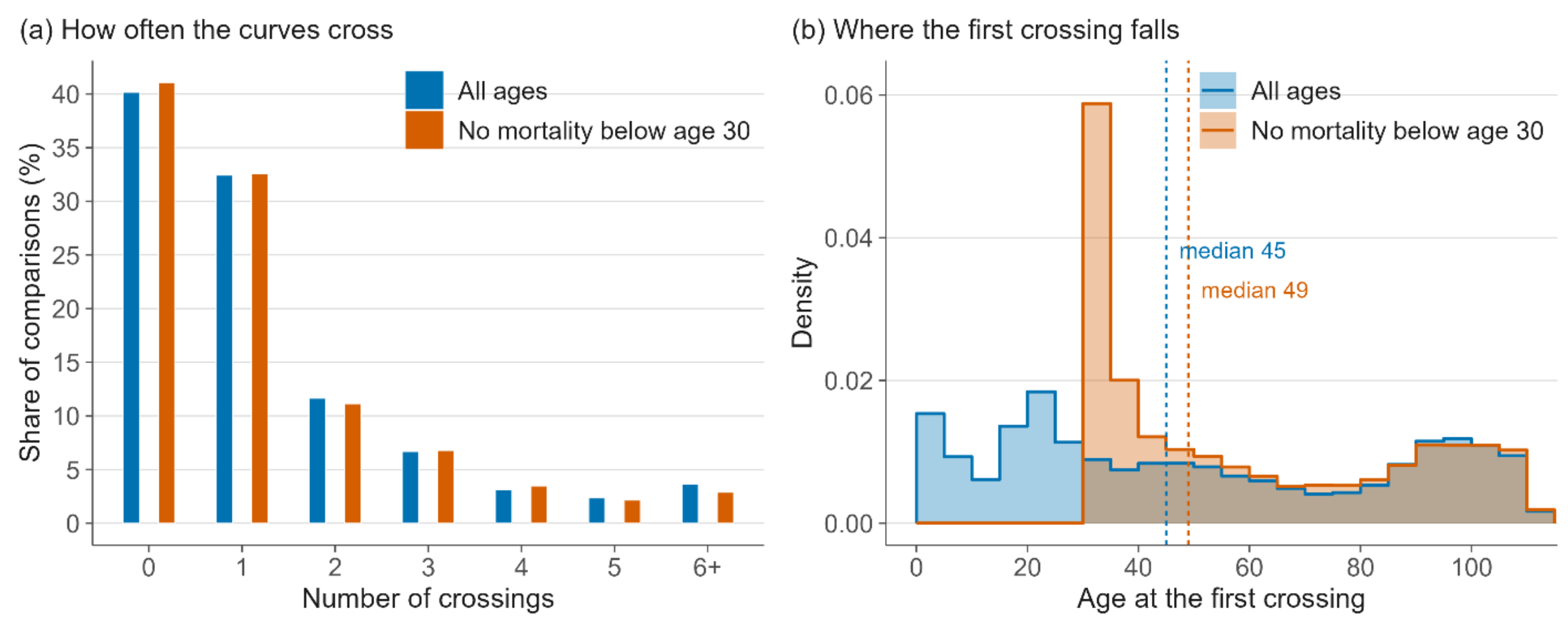

*Source: Own calculations from Human Mortality Database (2026).*

Assuming no mortality below age 30 sharpens the ordering (Table 2). It barely moves the cross-country figure (59.8% to 58.9%) and cuts its discrepancy from 0.118 to 0.080 years, because crossings below 30 occur where both curves are still close to one and the area between them is therefore tiny. For the sex gap the effect is large. Crossings fall from 10.2% to 4.3% and the median first crossing moves from age 12 to age 103. Accordingly, the small amount of crossing that remains between female and male survivorship sits almost entirely in the extreme tail.

## What crossings cost, and how much of the difference reverses

Crossings are common but usually very small in size. Across the cross-country pairs the mean $2\min(P,N)$ is 0.118 years against a mean $W_1$ of 3.69. Plotting $2\min(P,N)$ against $\Delta e_0$ reveals that highest gaps between $\Delta e_0$ and $W_1$ can be found for pairs with a rather small difference in life expectancy at birth (Figure 3). This is not surprising because these are the cases where a survival advantage of one population was later reversed, i.e., the difference in net survival is small but the two mortality regimes differ substantially. A large gap in $e_0$, however, refers to situations where a survival advantage for one population accumulates over age. Then, $\Delta e_0$ and $W_1$ are usually interchangeable.

Table 3 illustrates what the two measures add to $\Delta e_0$ on four individual comparisons. In the first, Japanese women survive American women at every age. One regime, $\Lambda = 0$, $\Psi = 1$, and $W_1 = \Delta e_0 = 5.84$ years exactly, so the life expectancy gap is a complete summary. The second is the case that demonstrates the relevance of weighting. Australian and Latvian men differ by 10.30 years, one of the largest gaps in the data, and their curves nonetheless cross at age 16, because Latvian child mortality in 2019 was marginally the lower. That regime is worth 0.005 years out of 10.31, so $\Lambda = 0.001$ and $\Psi = 1.001$ record correctly what the count cannot, that Australia dominates for every practical purpose.

**Fig. 3 The discrepancy $2\min(P, N)$ against the gap in life expectancy at birth, women and men, 1990 to 2020**

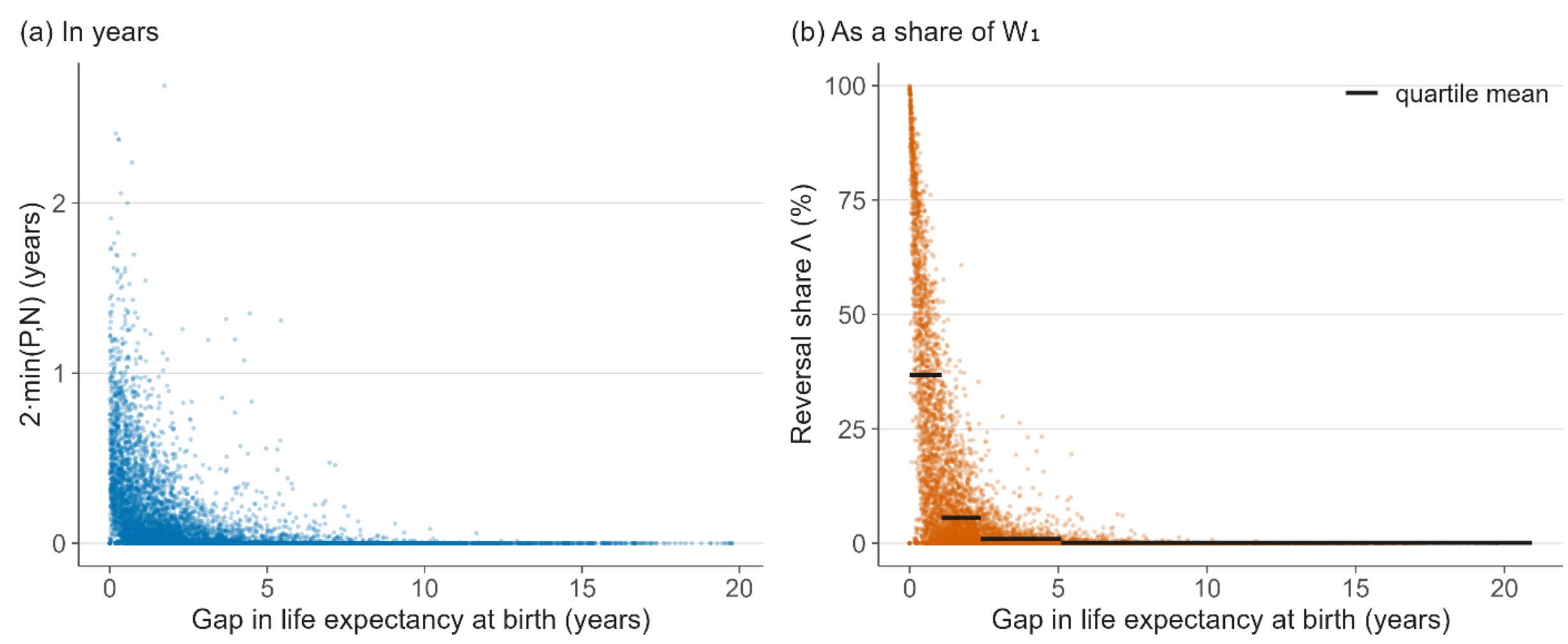

*Source: Own calculations from Human Mortality Database (2026).*

The third is the opposite. American and Czech men in 2019 had almost the same life expectancy, $\Delta e_0 = 0.27$ years, yet $W_1 = 2.63$. Czech men survived American men up to age 71, worth 1.18 years, and American men survived Czech men above it, worth 1.45. The two nearly cancel in terms of net survival but not in the absolute one, and $\Lambda = 0.899$ suggests that nine tenths of the distance between these populations is invisible to the gap in life expectancy. Reporting $\Delta e_0$ alone would suggest that the two populations are almost identical, when in fact they differ substantially. Finally, Spanish and Korean men are similar in terms of their $e_0$ level but their survival curves cross three times (at age 66, 80, and 102). The advantage regimes carry enough masses to be reflected in $\Psi$ (2.56).

**Table 3 Four comparisons in which $\Lambda$ and $\Psi$ add to $\Delta e_0$**

| Comparison | $\Delta e_0$ | $W_1$ | Cross. | Ages | Masses $m_r$ | $\Lambda$ | $\Psi$ |
|---|---|---|---|---|---|---|---|
| Japan vs. USA, women 2019 | 5.84 | 5.84 | 0 | — | 5.835 | 0.000 | 1.00 |
| Australia vs. Latvia, men 2019 | 10.30 | 10.31 | 1 | 16 | 0.005, 10.302 | 0.001 | 1.00 |
| USA vs. Czechia, men 2019 | 0.27 | 2.63 | 1 | 71 | 1.182, 1.448 | 0.899 | 1.98 |
| Spain vs. Korea, men 2019 | 0.40 | 0.57 | 3 | 66, 80, 102 | 0.207, 0.086, 0.279, 0.001 | 0.305 | 2.56 |

Own calculations from Human Mortality Database (2026).

## The correlation between the W1 Wasserstein distance and the non-overlap index as well as the Kullback -Leibler Divergence

The non-overlap index is strongly and monotonically related to $W_1$ (Pearson $r = 0.958$, Spearman $\rho = 0.958$), and the relationship is concave, flattening as $W_1$ grows, exactly as (20) implies. The pure-shift approximation is therefore useful as an explanation but not as a substitute. It over predicts the non-overlap index by 0.026 on average, with a mean absolute error of 0.046. The bias is in the expected direction and suggests that real mortality differences are not pure translations, where the shape of $d(x)$ also changes, the modal density $p(M)$ is no longer common to the two distributions and the first-order expansion overstates how much overlap is lost.

**Fig. 4 $W_1$ against two established dissimilarity measures**

*Source: Own calculations from Human Mortality Database (2026).*

The Kullback–Leibler divergence is likewise strongly associated with $W_1$ ($r = 0.936$, $\rho = 0.912$) but with a visibly convex and more dispersed relationship, consistent with its being a measure of information gain rather than of displacement. Both associations are high enough that the three measures would rank most pairs of populations similarly.

## How the population construction changes the crossings

Replacing the life table deaths by the observed ones on the same cross-country pairs changes the results substantially. The share of pairs whose survivorship curves cross rises from 59.8% to 78.4%, the share crossing more than once from 27.4% to 43.5%, and the mean discrepancy $2\min(P, N)$ nearly triples, from 0.118 to 0.298 years. The median first crossing moves from age 45 to age 59 (Table 4). The gap between the two constructions is stable across 1990 to 2020 and is present whether or not the youngest ages are included (Figure 5). Because both distributions rely on the same rates, this widening is entirely an effect of age structure: the observed populations are further from stationarity, and their age-at-death distributions cross more often as a result.

**Table 4 The three frameworks, on the cross-country pairs, women and men, 1990 to 2020**

| | | | Crossings | | | Years | | | Gap off |
|---|---|---|---|---|---|---|---|---|---|
| | Ages | $n$ | $\geq 1$ (%) | $> 1$ (%) | Median first age | Mean gap | Mean $2\min(P, N)$ | $r$ | by $> 1$ y (%) |
| *Complementary distribution functions* | | | | | | | | | |
| Life table | All | 69,188 | 59.8 | 27.4 | 45 | 3.57 | 0.118 | 0.998 | 1.5 |
| Constant birth | All | 5,440 | 70.7 | 37.5 | 34 | 1.70 | 0.170 | 0.980 | 1.7 |
| Population | All | 69,188 | 78.4 | 43.5 | 59 | 3.67 | 0.298 | 0.990 | 9.2 |
| Life table | 30+ | 69,188 | 58.9 | 26.4 | 49 | 3.25 | 0.080 | 0.998 | 1.2 |
| Constant birth | 30+ | 5,440 | 63.0 | 33.1 | 48 | 1.58 | 0.114 | 0.982 | 1.5 |
| Population | 30+ | 69,188 | 74.6 | 42.8 | 63 | 3.13 | 0.157 | 0.987 | 5.7 |

Own calculations from Human Mortality Database (2026).

**Fig. 5 Share of comparisons with at least one crossing of the framework's defining curve, by year and framework**

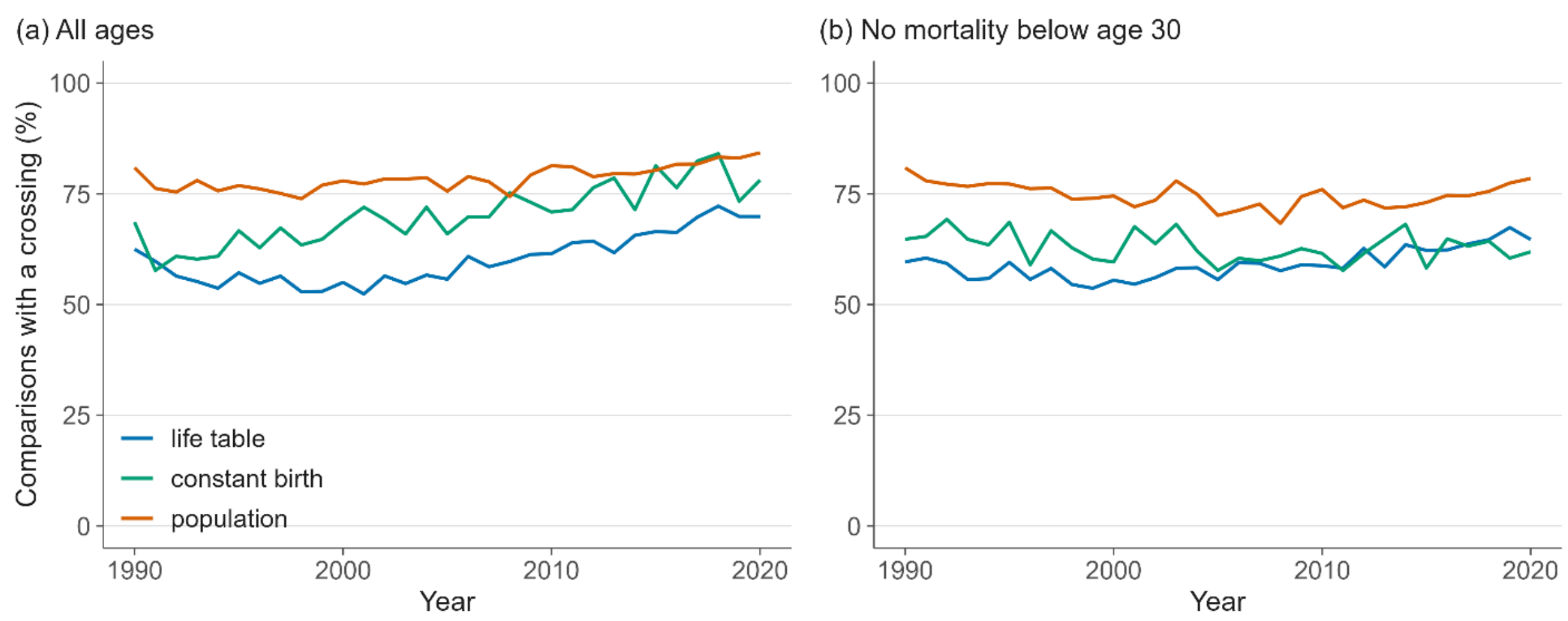

Source: Own calculations from Human Mortality Database (2026).

The number of crossings for the constant-birth population lies between the other two populations. The order holds for the term $2\min(P, N)$with 0.118 years in the life table, 0.170 in the constant-birth population, and 0.298 in the observed one. Yet, it should be mentioned that the constant-birth figures are based on the 5,440 comparisons available for the sixteen long-time series populations and not on the full 69,188 pairs data set.

## Decomposing the Wasserstein distance

Table 5 applies both decompositions to the two pairs of Figure 1. For Japan versus the United States in 2019, which is a pair with no crossing, the $W_1$ value is 5.84 and 3.44 of it (about 59%) can be contributed to shift, while 2.39 (about 41%) reflect dispersion. Using the $W_2$ measure and its decomposition instead the largest contribution stems from location (34.07 of 44.62 or about 76%). Accordingly, both frameworks agree that the distributions differ mainly in position. For England & Wales versus Iceland in 1849, where $\Delta e_0 \approx 0$, only about 4% of $W_1$ is a shift and 96% is dispersion, whereas $W_2$ splits between size (46%) and shape (54%) without a contribution for location. The two decompositions suggest therefore that this pair differs mainly in spread rather than position, but only the $W_2$ framework separates a difference in scale from a difference in shape.

Figure 6 applies the decompositions to mortality improvement within countries. The figures refer to mean $W_1$ and $W_2$ values across countries within a period, separated by sex. More specifically, we calculated the distances between two points in time for countries with available data (34 countries for 1960 to 1980, 35 countries for 1980 to 2000, and 37 countries for 2000 to 2019) and then simply averaged the values across countries. The pattern shows a clear trend. In 1960 to 1980, about half of the change in $W_1$ for women is a shift. For the period 1980 to 2000, the shift increases to 68% and by 2000 to 2019 to 78%. Among men, the increases in the shift contribution over time are even more pronounced as they jump from 43% in the first period to 81% in the last period.

**Table 5 Both decompositions for the two pairs of Figure 1.**

| | $W_1$ decomposition | | | $W_2$ decomposition | | | | |
|---|---|---|---|---|---|---|---|---|
| | $W_1$ | Shift | Disp. | $W_2$ | Location | Size | Shape | $\rho$ |
| Japan vs. United States, women, 2019 | 5.84 | 3.44 | 2.39 | 44.62 | 34.07 | 8.54 | 2.01 | 0.99 |
| England & Wales vs. Iceland, 1849 | 6.20 | 0.27 | 5.93 | 64.19 | 0.00 | 29.76 | 34.43 | 0.98 |

Source: Own calculations (Human Mortality Database 2026).

The squared distance tells the same story. For both women and men, the location share increases over time. This indicates that mortality improvements have become increasingly a translation of the whole age-at-death distribution to older ages rather than a compression of it.

**Fig. 6 Two Decomposition approaches of the Wasserstein distance**

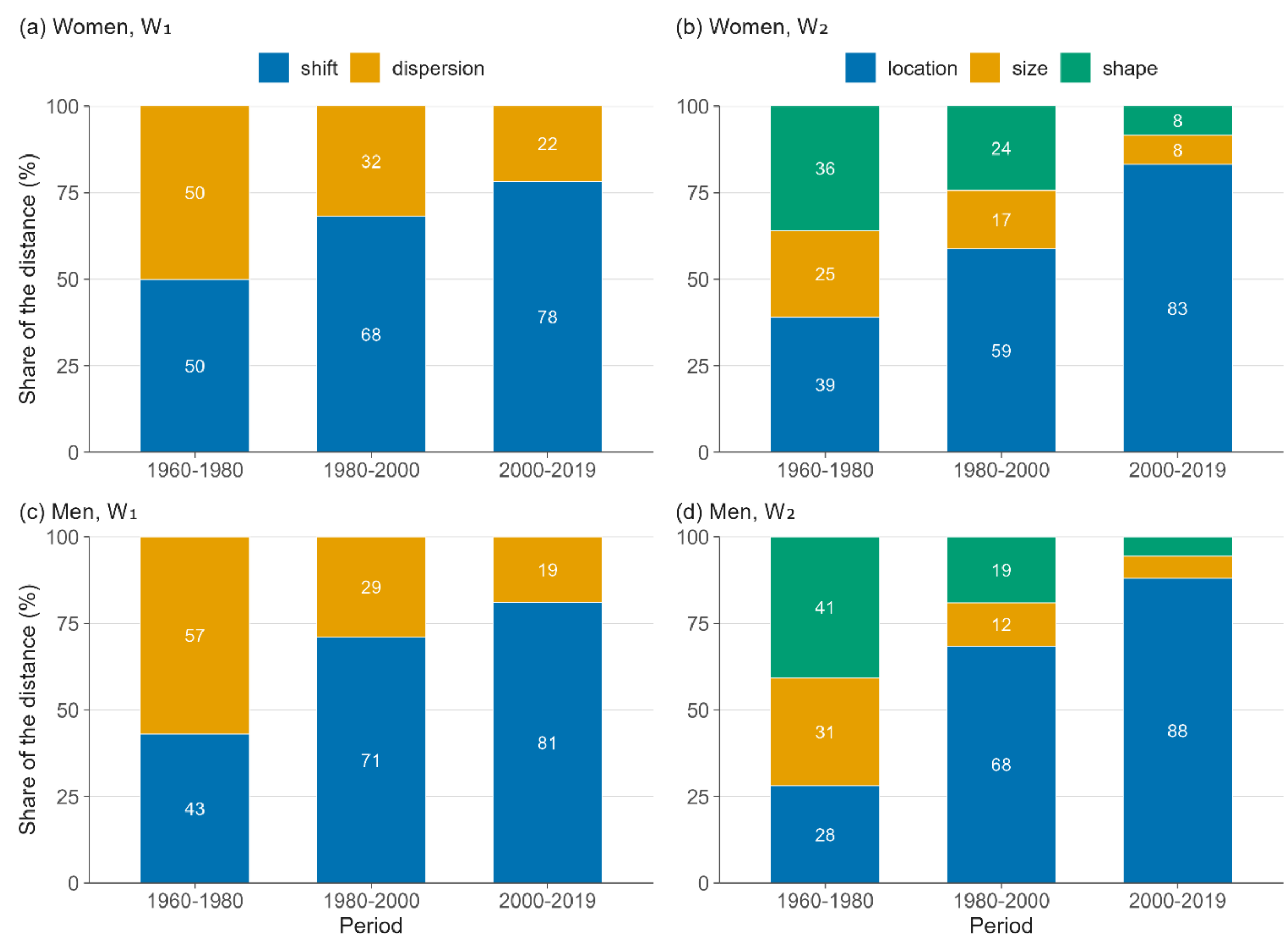

*Source: Own calculations from Human Mortality Database (2026).*

## Adding cause of death

Figure 7 depicts the cause-specific Wasserstein distance for France and the United States between the years 2015 to 2019. The cause-specific mortality data considers four categories, i.e., cancer, cardiovascular disease, external causes, and a residual category. The two-dimensional distance exceeds the one-dimensional one for every year and both sexes, and the gap widens with $\lambda$ as expected. This makes sense because adding a second dimension can only increase the transport cost. A larger $\lambda$ means that a higher cost for moving deaths between causes is chosen, leading to a higher Wasserstein distance by definition. We observe sex differences in the cause-specific Wasserstein distance time trend. Both distances are larger for women than for men, but adding the cause dimension increases the men's distance proportionally more, i.e., there are larger distances between the lines in Figure 7 among men. This indicates that French and American men differ more in what they die of, relative to how much they differ in when. Moreover, plotting the time trend for several choices of $\lambda$ demonstrates how sensitive the measure is to its value.

**Fig. 7 One- and two-dimensional Wasserstein distances between the cause-specific age-at-death distributions of France and the United States, 2015 to 2019**

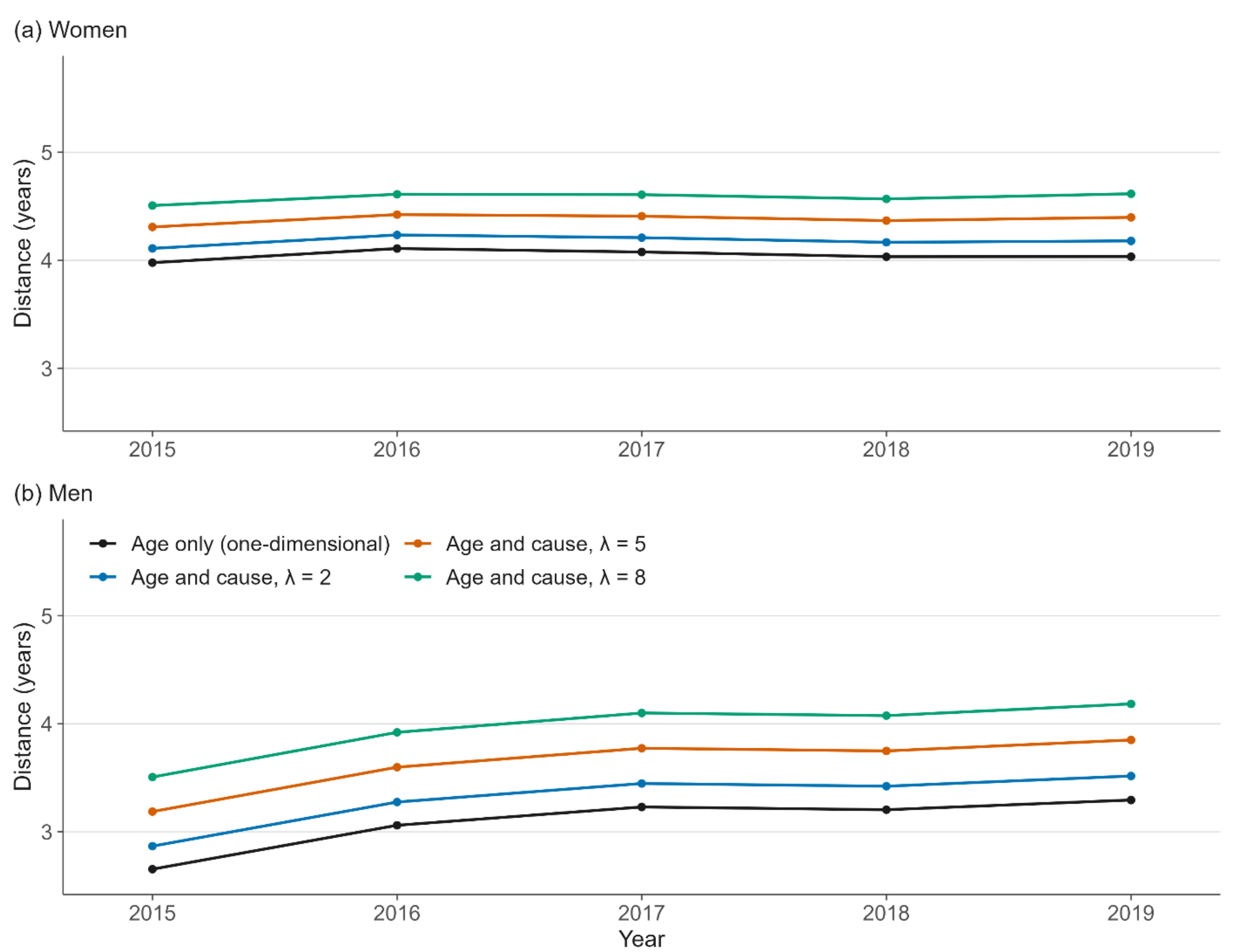

*Source: Own calculations from cause-of-death data (Human Mortality Database 2026).*

# Discussion

The difference in life expectancy at birth is a net difference in survival, whereas the Wasserstein distance $W_1$ reflects absolute difference in survival. In most comparisons the two agree closely and the residual $2\min(P,N)$ is small, suggesting that one population usually holds the survival advantage across the entire age range. This should not be interpreted as one population having lower death rate at every age. Instead, it is the accumulated survivorship advantage that does not change its sign, i.e., elevated death rates for some ages cannot reverse the previously accumulated survivorship advantage. In most cases in which we observe such a reversal the crossover occurs at very young or very old ages. In high-income countries, mortality is usually so low at young ages that the two curves may cross only due to fluctuation. For this reason, we also report a variant that assumes no mortality below age 30. Our empirical results indicate that excluding those ages leads indeed to a substantially lower number of observed crossings. At the oldest ages the exposures are small and the rates fluctuate as well. In addition, the survivorship functions differ only by quantities close to machine zero, so whether they are counted as crossing depends on the handling of floating-point arithmetic. Nevertheless, we deliberately decided not to truncate the oldest old (e.g., closing the life table at age 90+ or 100+) because this would move large share of the life table deaths into the open age interval. Especially for countries with a high life expectancy at birth value such as Japan or Hong Kong, this would substantially distort the shape of the age-at-death distribution.

It should be mentioned that our analysis concentrates on recent decades and on high-income countries. The crossing frequencies we report should be read as indicative rather than as final figures. A sample reaching further back in history, or into higher-mortality settings, may lead to a different picture of survivorship crossings. Further, the empirical analysis in this paper focuses on period life tables. Using cohort life tables might increase the share of pairs where mortality schedules between two populations differ more. Yet, the results provided by Sauerberg (2025) do not indicate that survivorship functions from cohort life tables cross substantially more often as compared to period life tables.

We introduce two measures, $\Lambda$ and $\Psi$, which can be reported along with $\Delta e_0$. They help identifying when two populations differ more in their mortality schedule than the life-expectancy gap alone would suggest. Moreover, we examine the impact of the underlying population model. Replacing the life table population by the observed population increases the crossing frequency substantially, while the constant-birth population produces a number of crossings that lies between the two. This comparison is interesting because all three constructions share the same set of period death rates and therefore, the increase in crossings can be attributed to differences in the age structure rather than to the mortality ratio. Among the three, the life table model is the most convenient for relating the difference in the means of two age-at-death distributions to $W_1$ because its survivorship function $l(x)$ is simultaneously the age structure of the stationary population and the complementary distribution function of $d(x)$.

In some settings a crossing cannot occur at all. Comparing a cause-eliminated life table with its all-cause counterpart necessarily produces a survivorship advantage at every age, and the same holds under any proportional-hazards model, such as the Gompertz model. In addition, every life expectancy at birth vs. health expectancy at birth comparison satisfies the no crossing assumption. In each of these cases condition (4) holds by construction, so $W_1$ reduces exactly to the difference in means and the residual $2\min(P, N)$ is zero.

The Wasserstein distance is strongly correlated with both the non-overlap index and the Kullback-Leibler divergence, even though each measure refers to a different concept of what makes two distributions differ, i.e., how far mass must move, how much mass is shared, and how much information separates them. For the non-overlap index we could find a formal relationship by expressing the non-overlap index as a function of $W_1$ under the assumption that the distributional difference is a pure shift along the age axis. Even though this assumption is not perfectly met in our empirical analysis, it may help explain the strong correlation between the two measures as well as identifying cases where the correlation is weaker. More importantly, the empirical analysis indicates that $W_1$ usually does not lead to different conclusions about distributional differences. The advantage of using $W_1$ over other measures is rather that it reduces to $\Delta e_0$ under a condition that can be checked. Further, the residual $2\min(P, N)$ is itself interpretable. It is therefore a natural addition to life expectancy differentials that can be easily reported along with them.

We also described two ways of decomposing the distance, which help to identify whether a difference between two age-at-death distributions is one of location or of shape. We would caution, however, against taking the Wasserstein distance approach for the narrower task of separating the compression of mortality from its shift to older ages, when what is wanted is a decomposition of the gap in $e_0$. The demographic methods of Bergeron-Boucher, Ebeling, and Canudas-Romo (2015) and of de Beer and Janssen (2016) are better suited to that purpose, since they do not rest on any assumption about whether the survivorship functions cross.

Finally, the extension to a second dimension, cause of death, is a mixed blessing. On the one hand, it is highly flexible. The cost of moving a death between causes can be set separately for each pair of causes, so that a transport from external causes to cancer can be made more costly than one from cancer to cardiovascular disease. But on the other hand, that same freedom is its weakness. The cost $\lambda$ has no natural scale, its choice is unavoidably somewhat arbitrary, and results computed under different choices are not readily comparable across studies.

# Conclusion

This paper introduced the Wasserstein distance to demography and linked it to the quantity the field relies on most, i.e., differences in $e_0$. When two survivorship functions do not cross, the $W_1$ between their age-at-death distributions is exactly the gap in life expectancy at birth. When they do cross, the two differ by the interpretable amount $2\min(P, N)$, so that the identity $W_1 = \Delta e_0 + 2\min(P, N)$ holds for every pair. Whether the curves cross is determined by the accumulated difference in age-specific death rates through the function $Z(x)$, from which we derived two magnitude-weighted summaries, the reversal share $\Lambda$ and the effective number of advantage regimes $\Psi$, that record how much of a mortality difference reverses with age and how that reversal is spread across it.

Empirically, survivorship crossings turn out to be ordinary rather than exceptional, but they are usually very small in size. Across our comparisons the discrepancy $2\min(P, N)$ rarely exceeds a fraction of a year, and it concentrates precisely where the life expectancy gap is small, which is the situation in which a crossover can reverse the ranking of two populations. It is therefore worthwhile examining the magnitude of crossings rather than simply counting the raw number of crossings.

Finally, the analysis indicates that the Wasserstein distance is strongly correlated with alternative inequality measures such as the non-overlap index or the Kullback-Leibler divergence. The Wasserstein distance framework leads to similar conclusions about differences in the age-at-death-distribution. Its advantage, however, is its direct connection to the gap in life expectancy at birth. Since it is expressed in the same unit of years, it can be reported alongside $\Delta e_0$ as a natural complement.

# Acknowledgements

**Data and code availability.** All data are publicly available from the Human Mortality Database (https://www.mortality.org) after free registration. The code is available at https://github.com/msauerberg/Wasserstein_repo. I used Claude by Anthropic to assist with the development of the code. AI was also used in the preparation of this paper, including brainstorming, improving the wording, translations, polishing equations, and formatting. The content was reviewed, verified, and finalized by me.

# Appendix

# Proof of Result 1

With a radix of one the life table age-at-death distribution is a probability density and $l(x)$ is the corresponding survival function, so the cumulative distribution function is $F(x) = 1 - l(x)$. Life expectancy at birth is the area under the survivorship function,

$$e_0 = \int_0^{\omega} l\,(x)\,dx, \qquad \text{hence} \qquad e_{0,A} - e_{0,B} = \int_0^{\omega} [l_A(x) - l_B(x)]\,dx. \qquad (24)$$

From (3), the Wasserstein distance is $W_1(d_A, d_B) = \int_0^{\omega} |\,F_A(x) - F_B(x)|\,dx$, and since

$$|F_A(x) - F_B(x)| = |(1 - l_A(x)) - (1 - l_B(x))| = |l_A(x) - l_B(x)|, \qquad (25)$$

condition (4) makes the integrand non-negative, so the absolute value may be removed:

$$W_1(d_A, d_B) = \int_0^{\omega} |\,l_A(x) - l_B(x)|\,dx = \int_0^{\omega} [l_A(x) - l_B(x)]\,dx. \qquad (26)$$

Comparing with (24) gives $W_1 = e_{0,A} - e_{0,B}$. □

**Life expectancy as the mean age at death.** Integrating (24) by parts,

$$\int_0^\omega l\,(x)\,dx = [x\,l(x)]_0^\omega - \int_0^\omega x\;l'(x)\,dx = \int_0^\omega x\,[-l'(x)]dx, \qquad (27)$$

because $l(\omega) = 0$. Since $d(x) = -l'(x)$ by definition and $\int_0^\omega d\,(x)\,dx = 1$ for a radix of one,

$$e_0 = \int_0^\omega x\;d(x)\,dx = \frac{\int_0^\omega x\;d(x)\,dx}{\int_0^\omega d\,(x)\,dx}, \qquad (28)$$

the second form holding for any radix. Life expectancy at birth is therefore both the area under the survivorship function and the mean of the age-at-death distribution, so Result 1 can be restated as $W_1(d_A, d_B) = \bar{d}_A - \bar{d}_B$ whenever (4) holds.

# Relation to the Non-Overlap Index

Moulton and Jiang (2018) show that the Jaccard index of two probability densities is an exact function of the total variation distance,

$$J = \frac{1 - \mathrm{TV}}{1 + \mathrm{TV}}, \qquad \mathrm{TV}(p, q) = \frac{1}{2}\int_{-\infty}^{\infty} |\,p(x) - q(x)|\,dx, \qquad (29)$$

so $J$ and TV carry the same information. To connect TV to $W_1$, consider a pure location shift $q(x) = p(x - \delta)$ with $\delta > 0$. Then $\mathrm{TV}(\delta) = 1/2\int|p(x) - p(x - \delta)|\,dx$, and a first-order Taylor expansion $p(x - \delta) \approx p(x) - \delta p'(x)$ gives

$$\mathrm{TV}(\delta) \approx \frac{\delta}{2}\int_{-\infty}^{\infty} |\,p'(x)|\,dx. \qquad (30)$$

For a unimodal density $p'$ changes sign exactly once, at the modal age $M$. The rising part contributes $\int_{-\infty}^{M} p'\,(x)\,dx = p(M)$ and the falling part $\int_M^\infty -p'(x)\,dx = p(M)$, so $\int|p'(x)|\,dx = 2p(M)$ and

$$\mathrm{TV}(\delta) \approx \delta\,p(M). \qquad (31)$$

Under a pure shift the Wasserstein distance is exactly the shift, $W_1 = \delta$. Combining this with (31) and the exact relation (29) yields

$$J \approx \frac{1 - W_1\,p(M)}{1 + W_1\,p(M)}, \qquad \text{equivalently} \qquad 1 - J \approx \frac{2\,W_1\,p(M)}{1 + W_1\,p(M)}, \qquad (32)$$

which is (20). The non-overlap index is thus a function of $W_1$ alone up to the scaling factor $p(M)$, and the relationship is a strictly concave increasing curve that flattens as $W_1$ grows, because the denominator increases with $W_1$. In the empirical implementation we take $p(M)$ to be the mean of the two modal densities of the pair.

# The Two Decompositions

## Shift and dispersion

Resin et al. (2025) express $W_1$ through quantile functions,

$$W_1(P,Q) = \int_0^1 \left|F_P^{-1}(u) - F_Q^{-1}(u)\right| du = \frac{1}{2}\int_0^1 \text{AVM}_\alpha\,(P,Q)\, d\alpha, \qquad (33)$$

with

$$\text{AVM}_\alpha(P,Q) = \left|F_P^{-1}(1+\alpha/2) - F_Q^{-1}(1+\alpha/2)\right| + \left|F_P^{-1}(1-\alpha/2) - F_Q^{-1}(1-\alpha/2)\right|. \qquad (34)$$

The pointwise components are

$$\text{Shift}_{\alpha,+}^{W_1}(P,Q) := 2\left[\min\left\{F_p^{-1}\left(1+\frac{\alpha}{2}\right) - F_Q^{-1}\left(1+\frac{\alpha}{2}\right), F_p^{-1}\left(1-\frac{\alpha}{2}\right) - F_Q^{-1}\left(1-\frac{\alpha}{2}\right)\right\}\right]_+, \qquad (35)$$

$$\text{Disp}_{\alpha,+}^{W_1}(P,Q) := \left[\left(\mathrm{F}_\mathrm{P}^{-1}\left(1+\frac{\alpha}{2}\right) - \mathrm{F}_\mathrm{Q}^{-1}\left(1+\frac{\alpha}{2}\right) - \mathrm{F}_\mathrm{P}^{-1}\left(1-\frac{\alpha}{2}\right) - \mathrm{F}_\mathrm{Q}^{-1}\left(1-\frac{\alpha}{2}\right)\right)\right]_+, \qquad (36)$$

with the negative counterparts defined by exchanging $P$ and $Q$, $\text{Shift}_-^{W_1}(P,Q) := \text{Shift}_+^{W_1}(Q,P)$ and $\text{Disp}_-^{W_1}(P,Q) := \text{Disp}_+^{W_1}(Q,P)$. Integrating as in (33) gives the four components, and

$$W_1 = \text{Shift}_+^{W_1} + \text{Shift}_-^{W_1} + \text{Disp}_+^{W_1} + \text{Disp}_-^{W_1}. \qquad (37)$$

In the main text we report the totals $\text{Shift} = \text{Shift}_+ + \text{Shift}_-$ and $\text{Disp} = \text{Disp}_+ + \text{Disp}_-$. See Resin et al. (2025) for more details. The implementation of their decomposition method in R can be found here: https://github.com/resinj/replication_SD-Decomp.

## Location, size, and shape

For the squared second-order distance, expanding $\int_0^1 (F_A^{-1} - F_B^{-1})^2 du$ and collecting terms gives (23): writing $\mu$ and $\sigma$ for the mean and standard deviation of each age-at-death distribution and $\rho$ for the Pearson correlation between the quantile functions,

$$W_2 = (\mu_A - \mu_B)^2 + (\sigma_A - \sigma_B)^2 + 2\sigma_A\sigma_B(1-\rho). \qquad (38)$$

See Schefzik et al. (2021) for more details.